\documentclass[showpacs,floatfix,citeautoscript,twocolumn,prb,superscriptaddress]{revtex4-1}

\usepackage{graphicx,xcolor}
\usepackage{amsmath}
\usepackage{comment}
\usepackage{textcomp}
\usepackage{inputenc}

\newcommand{\kk}[1]{\textcolor{olive}{{\bf Kisung: #1 }}}

\begin{document}

\title{In-plane magnetic structure and exchange interactions in the high-temperature antiferromagnet Cr$_2$Al}

\author{Chengxi Zhao}
\author{Kisung Kang}
\affiliation{Department of Materials Science and Engineering and Materials Research Laboratory, University of Illinois at Urbana-Champaign, Urbana, IL 61801, USA}
\author{Joerg C. Neuefeind}
\affiliation{Neutron Scattering Division, Oak Ridge National Laboratory, Oak Ridge, TN 37831, USA}
\author{Andr\'{e} Schleife}
\affiliation{Department of Materials Science and Engineering and Materials Research Laboratory, University of Illinois at Urbana-Champaign, Urbana, IL 61801, USA}
\affiliation{National Center for Supercomputing Applications, University of Illinois at Urbana-Champaign, Urbana, IL 61801, USA}
\author{Daniel P. Shoemaker}\email{dpshoema@illinois.edu}
\affiliation{Department of Materials Science and Engineering and Materials Research Laboratory, University of Illinois at Urbana-Champaign, Urbana, IL 61801, USA}


\begin{abstract}

The ordered tetragonal intermetallic Cr$_2$Al forms the same structure type as Mn$_2$Au, and the latter has been heavily investigated for its potential in antiferromagnetic spintronics due to its degenerate in-plane N\'{e}el vector. We present the single crystal flux growth of Cr$_2$Al and orientation-dependent magnetic properties. Powder neutron diffraction of Cr$_2$Al and first-principles simulations reveal that the magnetic ordering is likely in-plane and therefore identical to Mn$_2$Au, providing a second material candidate in the MoSi$_2$ structure type to evaluate the fundamental interactions that govern spintronic effects.  The single ordering transition seen in thermal analysis and resistivity indicates that no canting of the moments along the $c$ axis is likely.  Magnetometry, resistivity, and differential scanning calorimetry measurements confirm the N\'{e}el temperature to be $634 \pm 2$~K. First-principles simulations indicate that the system has a small density of states at the Fermi energy and confirm the lowest-energy magnetic ground state ordering, while Monte Carlo simulations match the experimental N\'{e}el temperature.

\end{abstract}

\maketitle 

\section{Introduction} 

Antiferromagnets have attracted considerable attention due to fundamentally different proposed magnetization switching mechanisms and optically-probed spin relaxation dynamics in the THz range,\cite{Kimel,Little,Kamp} compared to GHz for ferromagnets and ferrimagnets.\cite{Viala,Lee,Chai,Sharma}
Recent studies have demonstrated partial switching of the N\'{e}el vector of antiferromagnetic CuMnAs and Mn$_2$Au\cite{Bhat2018,Barthem2013,Jourdan2015}, perhaps due to current-induced spin-orbit torques.\cite{Olejnik,Saidl,Wadley2015,Wadley2016,Wadley2018} The magnetic-field-induced spin rotation and spin-flop behavior of epitaxial CuMnAs thin films has been probed using X-ray magnetic linear dichroism and signatures of the N\'{e}el vector reorientation have been seen in the anisotropic magnetoresistance with multiterminal devices.\cite{Wang,Wadley2016,Olejnik}  N\'{e}el vector reorientations have also been observed in femtosecond pump-probe magneto-optical Kerr effect experiments.\cite{Saidl}

Intriguingly, Chien \emph{et al.}\ have shown that the purported spin-orbit torque magnetoresistance effects may be artifacts of excessive current-induced heating through multiterminal devices, thus unequivocal detection of the N\'{e}el vector before and after SOT switching at low currents is required to effectively prove magnetic switching.\cite{Chien}
The in-plane current-pules switching experiments on Mn$_2$Au quantified the anisotropic magnetoresistance and planar Hall effect of thin films.\cite{Bodnar}  A spin-orbit-torque-driven antiferromagnetic resonance of Mn$_2$Au has not been observed to date by time-domain THz spectroscopy, despite early reports.\cite{bhattacharya_retraction_2019}

Mn$_2$Au remains antiferromagnetic until it forms a disordered (Mn,Au) solid solution at 953~K,\cite{Cahn} with a N\'{e}el temperature predicted to be above 1600~K, due to its large anisotropy energy, calculated based on first-principles calculations.\cite{Khme}
Such large anisotropy may not be desired in all spintronic applications. In particular, the ability to observe a N\'{e}el-order-driven effect vanish at $T_N$, without excessive current flow and Joule heating, would be a strong confirmation of true spintronic effects. To understand the underlying physics and capabilities, there is a pressing need to expand the library of intermetallic antiferromagnets that share features with CuMnAs (which itself has not been grown as bulk single crystals, likely due to a competing orthorhombic phase)\cite{Uhlirova} and Mn$_2$Au. Single crystals of such materials are needed to study the orientation dependence of their magnetic dynamics.

\begin{figure}
\centering\includegraphics[width=0.9\columnwidth]
{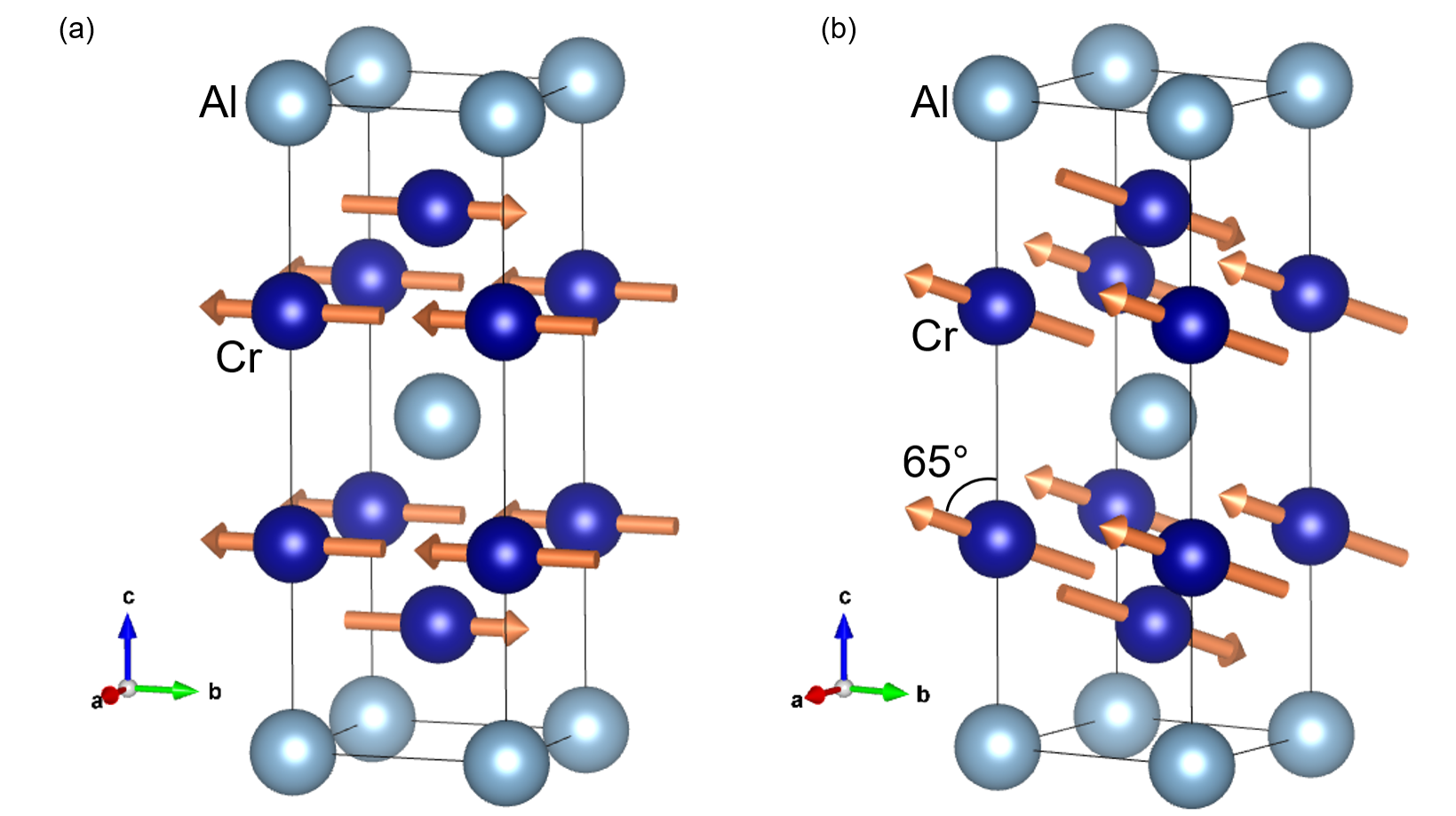} \\
\caption{\label{fig:structure}
The $k$=0 magnetic structure of Cr$_2$Al is shown in (a), with Al light and Cr dark. The reported magnetic structure from Atoji in (b) has a 65$^\circ$ canting away from the $c$ axis.\cite{Atoji} No data in our study point to a reduction in symmetry from our in-plane model to that of Atoji.
}
\end{figure}

Cr$_2$Al is an incongruently-melting intermetallic antiferromagnet with the MoSi$_2$ structure type, tetragonal space group $I4/mmm$, isostructural to Mn$_2$Au and Au$_2$Mn. There is no reported single crystal growth or measurements of its anisotropic properties. Atoji presented variable-temperature powder neutron diffraction of Cr$_2$Al in 1965. \cite{Atoji} 
In that study, the (001) neutron diffraction peak intensity was used to determine the $T_N = 598 \pm 5$~K. The magnetic structure was claimed to be inclined by an angle $65 \pm 2.4^\circ$ toward the $c$ axis, though no explanation of that determination was given. We show here that the in-plane (zero canting) spin configuration is far more plausible.

In 1988, Semukhin \emph{et al.}\ investigated the magnetic phase transition of Cr$_2$Al by high-temperature X-ray diffraction and calorimetry.\cite{Semukhin} Sharp changes in the $a$ and $c$ axis lattice parameters and a heat capacity anomaly occurred $655 \pm 5$~K. Variation in the reported $T_N$ may be due to a compositional width of the Cr$_2$Al phase, which has been reported to be about 10\,\%.\cite{Mahdouk,rank_experimental_2019} Variable $T_N$ can also arise from further alloying, as when Susner \emph{et al.}\ showed that the resistivity anomaly at $T_N$ shifts to lower temperatures and is more pronounced with increased Fe substitution for Cr.\cite{Susner} 

Here we show the first steps of understanding the potential of Cr$_2$Al as a spintronic material, with single crystal growth via tin flux and melt decanting. 
We consider the allowed magnetic orderings and examine the sensitivity to a $c$-axis component of the local moments.  Magnetometry measurements with varying temperature and external field  show the predicted magnetic susceptibility and anisotropy, while resistivity and calorimetry confirm the presence of a single $T_N$, without a second metamagnetic transition that would be expected in the case of spin canting as proposed by Atoji.\cite{Atoji} First-principles simulations test multiple magnetic configurations, with the experimental ordering found to be the predicted ground state. Exchange coefficients and band structures show how Cr$_2$Al has a small density of states near the Fermi energy, and confirm the experimental susceptibility and $T_N$.

\section{Methods}

Tetragonal single crystals with edge lengths around 0.5~mm and mass $\sim$1-2~mg were grown from Sn flux. 
Cr ($>$99.99\% metals basis) and Al (99.9999\% metals basis) powders were mixed in 2:1.5 molar ratio along with 95 at\% Sn powder (99.98\% metals basis) inside an Ar filled glove box and vacuum sealed inside a 13 mm inner diameter quartz tube. A stainless-steel mesh was inserted inside the tube just above the powder mixture.
The tube was heated to 1173~K at 10~K/min and held for 12 hours, slowly cooled to 933~K at 2~K/hr, and then the tube was taken out, flipped, and centrifuged to filter the grown crystals with the inserted mesh, followed by an air quench to room temperature. The collected crystals were sonicated in 1~M nitric acid for 2 hours to remove residual Sn.
Separately, Cr$_2$Al powder was prepared by annealing a mixture of Cr and Al powders with molar ratio of 2:1 that was pressed into a pellet and heated at 1073~K for 72 hours.

Scanning electron microscopy was performed in a JEOL 6060 LV SEM, and X-ray diffraction was performed in a Bruker D8 ADVANCE diffractometer with Mo K$\alpha$ radiation.
Powder neutron diffraction was performed on the NOMAD instrument at the Spallation Neutron Source at Oak Ridge National Laboratory.\cite{Calder}
Nuclear and magnetic structures were refined using GSAS-II\cite{Toby} and structures were visualized using VESTA.\cite{Momma}
After alignment by XRD, single crystal magnetometry was conducted using a standard quartz rod (2 to 400~K) and oven attachment (up to 750~K) on a Quantum Design MPMS3 vibrating sample magnetometer. 
High-temperature resistivity up to 773~K was measuring using a two-point pressed contact configuration shown in Figure S1.\cite{supplement} A separate thermocouple was placed on the sample stage to measure the temperature and the measurement was performed under flowing nitrogen atmosphere.
Differential scanning calorimetry was performed with 2.5 mg of crystals on a TA DSC 2500.

First-principles density functional theory (DFT) simulations were performed using the Vienna \emph{Ab-Initio} Simulation Package \cite{Kresse:1996,Kresse:1999} (VASP). In solving the Kohn-Sham equation, we used the generalized-gradient approximation (GGA) formulated by Perdew, Burke, and Ernzerhof \cite{Perdew:1997}, to describe exchange and correlation. The electron-ion interaction was described by the projector-augmented wave \cite{Blochl:1994} (PAW) method. For the Brillouin zone sampling, $21\,\times\,21\,\times\,7$ Monkhorst-Pack (MP) \cite{Monkhorst:1976} $\mathbf{k}$-points were used.
The kinetic energy cutoff of the plane-wave basis was chosen as 600~eV by convergence testing. Phonon dispersion calculations within the finite displacement method were performed using the \textsc{phonopy} package \cite{Phonopy:2015}, a supercell of $3\,\times\,3\,\times\,1$, and a $4\,\times\,4\,\times\,4$ MP $\mathbf{k}$-point grid. We performed all energy dispersion calculations accounting for noncollinear magnetism and spin-orbit coupling \cite{Steiner2016}.

Exchange coefficients are calculated using the spin polarized relativistic Korringa-Kohn-Rostoker (\texttt{SPR-KKR}) code \cite{Ebert:2011} and the relaxed atomic structure from the DFT calculations described above.
To integrate over the Brillouin zone, 1000 randomly chosen $k$-points are used in our energy convergence test with a criterion of $0.01$ meV/atom. Next, the exchange coefficients are extracted by Lichtenstein's approach, as implemented in the \texttt{SPR-KKR} code \cite{Liechtenstein:1984}. The exchange coefficients follow from the isotropic exchange term in a Heisenberg model,
\begin{equation}
\label{eq:exchange}
\mathcal{H}_{ex} = - \sum_{i \neq j}J_{ij}e_{i}e_{j}.
\end{equation}
The exchange Hamiltonian ($\mathcal{H}_{ex}$) consists of isotropic exchange coefficients ($J_{ij}$) and the unit vector of all magnetic moments at site $i$ and $j$ ($e_i$ and $e_j$). $J_{ij}$ is calculated up to an relative interaction distance $d=4.0a$, where $a$ is the $a$ axis lattice parameter of Cr$_2$Al.

We calculate atomistic spin dynamics using a Monte Carlo approach implemented in \texttt{UppASD}\cite{Eriksson:2017} to estimate the N\'eel temperature. To simulate microscopic magnetism at finite temperature, the stochastic Landau-Lifshitz-Gilbert (LLG) equation\cite{Eriksson:2017} is solved,
\begin{equation}
\label{eq:StochasticLLG}
\begin{split}
\frac{d\mathbf{m}_{i}}{dt} = &- \gamma_{L}\mathbf{m}_{i} \times (\mathbf{B}_{i}+\mathbf{B}_{i}^\mathrm{fl})\\ 
&- \gamma_{L} \frac{\alpha}{m_{i}} \mathbf{m}_{i} \times \left[ \mathbf{m}_{i} \times (\mathbf{B}_{i}+\mathbf{B}_{i}^\mathrm{fl}) \right],
\end{split}
\end{equation}
where $\alpha$ is an isotropic Gilbert damping constant and $\gamma_{L}=\gamma / (1+\alpha^2)$ is the renormalized gyromagnetic ratio. $\mathbf{m}_{i}$ is the magnetic moment at site {$i$} and here we use the value from our DFT ground state simulations.
$\mathbf{B}_{i}$ is the effective magnetic field containing exchange, anisotropy, and magnetic dipolar interactions at magnetic site $i$. The stochastic term is introduced by the thermally fluctuating magnetic field term $\mathbf{B}_{i}^\mathrm{fl}$. This term is shaped based on a central limit theorem as a form of Gaussian distribution with zero mean and a variance that depends on temperature. All atomistic spin dynamics calculations use a $12 \times 12 \times 4$ supercell.

\section{Results and Discussion}

\begin{figure}
\centering\includegraphics[width=0.9\columnwidth]
{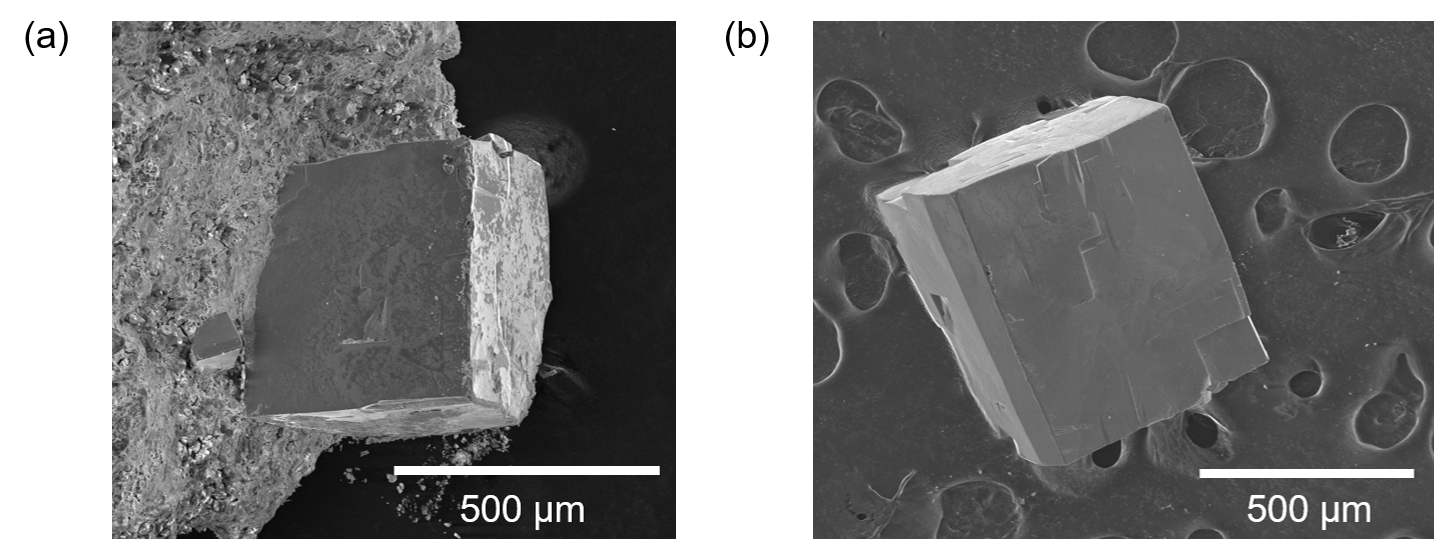} \\
\caption{\label{fig:SEM}
SEM micrographs of Cr$_2$Al crystals grown from Sn flux, after centrifuging and (a) before and (b) after removal of Sn with 1 M nitric acid.
}
\end{figure}

\begin{figure}
\centering\includegraphics[width=0.9\columnwidth]{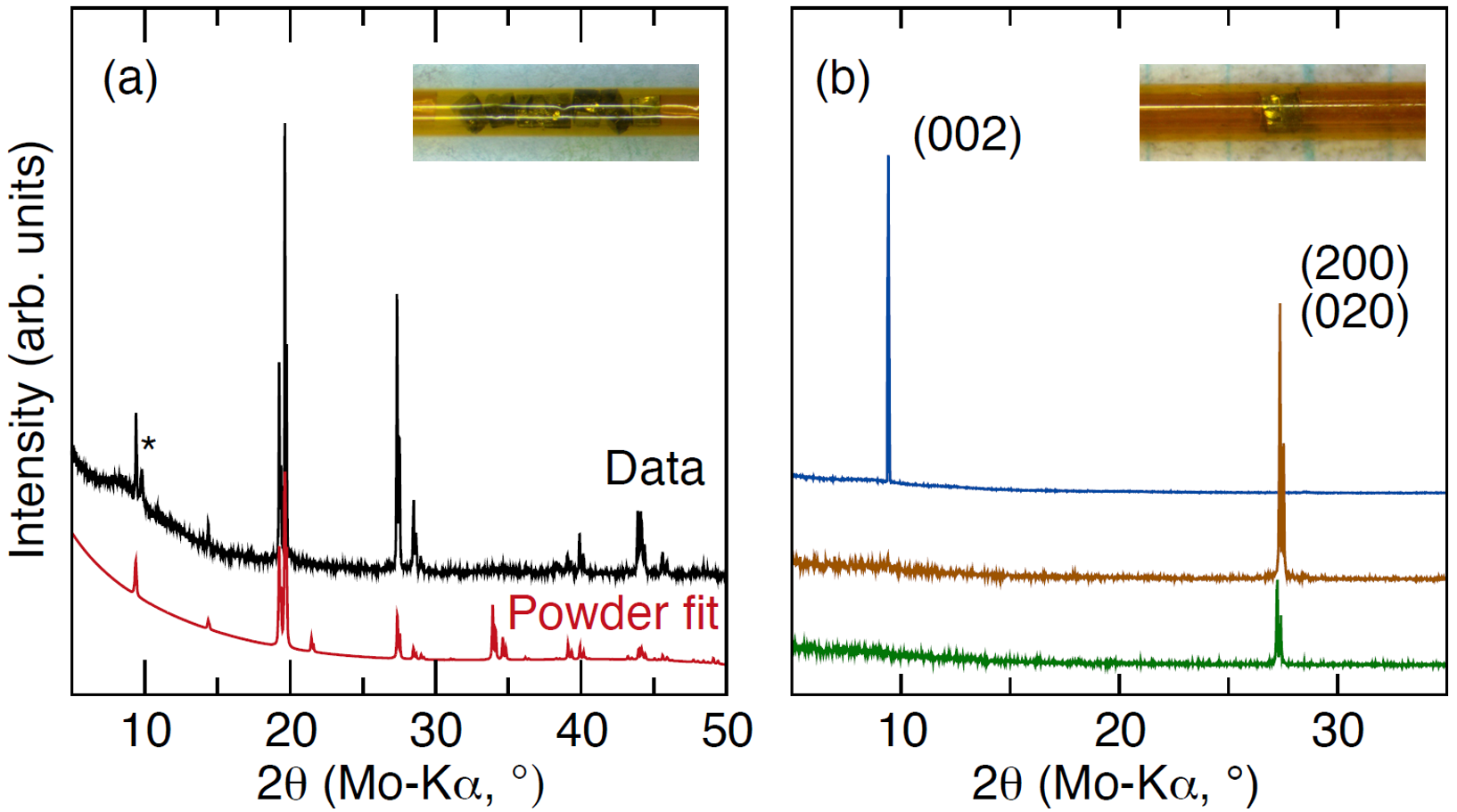} \\
\caption{\label{fig:xrd}
X-ray diffraction data for Cr$_2$Al crystals are shown for (a) multiple crystals collected simultaneously, along with a Rietveld-refined model for a powder. The asterisked peak is from the paraffin wax mount. In (b), data are shown for a single crystal oriented in three orthogonal directions and collected with the sample stationary. Insets show the measurement configurations with crystals loaded in kapton capillaries.
}
\end{figure}

The crystals obtained from melt centrifuging were metallic silver and had rectangular prismatic shapes, with a typical appearance shown in Figure \ref{fig:SEM}, before and after etching excess Sn in nitric acid. The phase purity was confirmed by powder XRD of several aligned crystals in a capillary and measured while rotating, which gives peaks that more fully reproduce the powder pattern of Cr$_2$Al in Figure \ref{fig:xrd}(a), with some expected difference in peak intensity due to preferred orientation. Collection of XRD data on stationary single crystals was used to determine the direction of the $c$ axis, with XRD shown for a typical crystal measured on three orientations shown in Figure \ref{fig:xrd}(b). Energy-dispersive X-ray spectroscopy gave an average Cr:Al ratio of 2.0(2), shown in Figure S2.\cite{supplement}

\begin{figure*}
\centering\includegraphics[width=0.8\paperwidth]{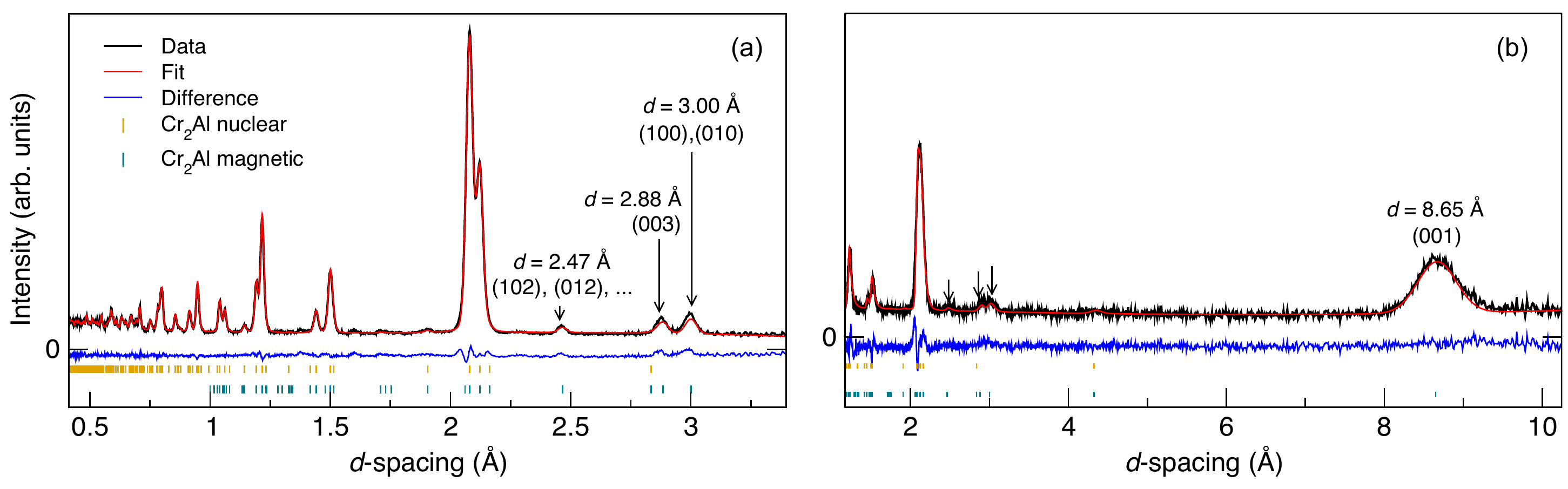} \\
\caption{\label{fig:NOMAD}
Time-of-flight neutron diffraction patterns and Rietveld-refined patterns from NOMAD detector banks with average collection angles $2\theta = 65^\circ$ (a) and $2\theta = 15.1^\circ$ (b). The two banks are optimized for collection on different ranges of $d$-spacings. Blue and red tick marks indicate nuclear and magnetic peaks, respectively. Magnetic peaks are indicated by arrows and indexing.
}
\end{figure*}

Figures \ref{fig:NOMAD}(a) and \ref{fig:NOMAD}(d) show the neutron diffraction pattern of a Cr$_2$Al powder at 298~K. The nuclear peaks give a clear fit to the ordered $I4/mmm$ structure, with the magnetic peak (100) clearly visible at $d=8.65$~\AA. The magnetic propagation vector $k = (0 0 0)$ reproduces all magnetic peaks, which confirms the magnetic unit cell exhibits the same dimension as the crystallographic unit cell. The magnetic space group of Cr$_2$Al with the N\'eel vector along [100] is $P_{I}nnm$ in Belov-Neronova-Smirnova (BNS) notation and $I_Pm'm'm$ in Opechowski-Guccione (OG) notation, and the refinement in Figure \ref{fig:NOMAD} is thus constrained with Cr1 moments on the body center sites opposite to the Cr2 moments on the edges of the cell. 

The refined magnetic moments on Cr positions are 1.06(19)$\mu_B$. The obtained magnetic structure resolved is visualized in Figure \ref{fig:structure}.
Powder diffraction is not sensitive to the orientation of the Cr moments within the $ab$ plane. The refined structure, with the nearest-neighbor Cr spins antiferromagnetic, which are in turn ferromagnetic across the Al plane, gives a significantly improved fit over a model where the Cr bilayers have ferromagnetic alignment within the layers, or antiferromagnetic alignment from layer to layer. These fits to the neutron diffraction data are shown Figures S4-5.\cite{supplement} The latter configurations relax to a nonmagnetic state in first-principles simulations. Likewise, a uniaxial configuration with spins along $c$ does not reproduce the neutron data (Figure S6-8).\cite{supplement} A small canting of the moments toward $c$, as suggested by Atoji,\cite{Atoji} does not significantly improve the fit, even with the addition of an extra free parameter, shown in Figure S9.\cite{supplement}
Small canting is, thus, unlikely due to the additional irreducible representation that would be required to observe it, along with the presence of a single N\'{e}el transition, which we will discuss subsequently. Canting the moment along $c$ also leads to a calculated increase in the total energy, as shown in Figure S10.

First-principles density functional theory simulations for antiferromagnetic Cr$_2$Al are implemented for ground state calculations, electronic band structure, phonon dispersion, magnetic susceptibility, and exchange coefficients. The calculated ground state confirms the $I4/mmm$ structure with $a=2.98$ \AA\ and $c=8.63$ \AA, in good agreement with the room temperature lattice parameters $a=3.00$ \AA\ and $c=8.65$ \AA\ from neutron diffraction in Figure \ref{fig:NOMAD}. Magnetic moments on Cr sites shown in Figure\,\ref{fig:structure} converged to 1.311\,$\mu_{B}$. The discrepancy of 23\,\% likely arises from decrease of the low-temperature moment upon heating to 300~K. Ground state DFT simulations for other possible magnetic orderings (see Figure S4-5)\cite{supplement} conclude that there is no stable state for those.


\begin{figure}
\centering\includegraphics[width=0.9\columnwidth]{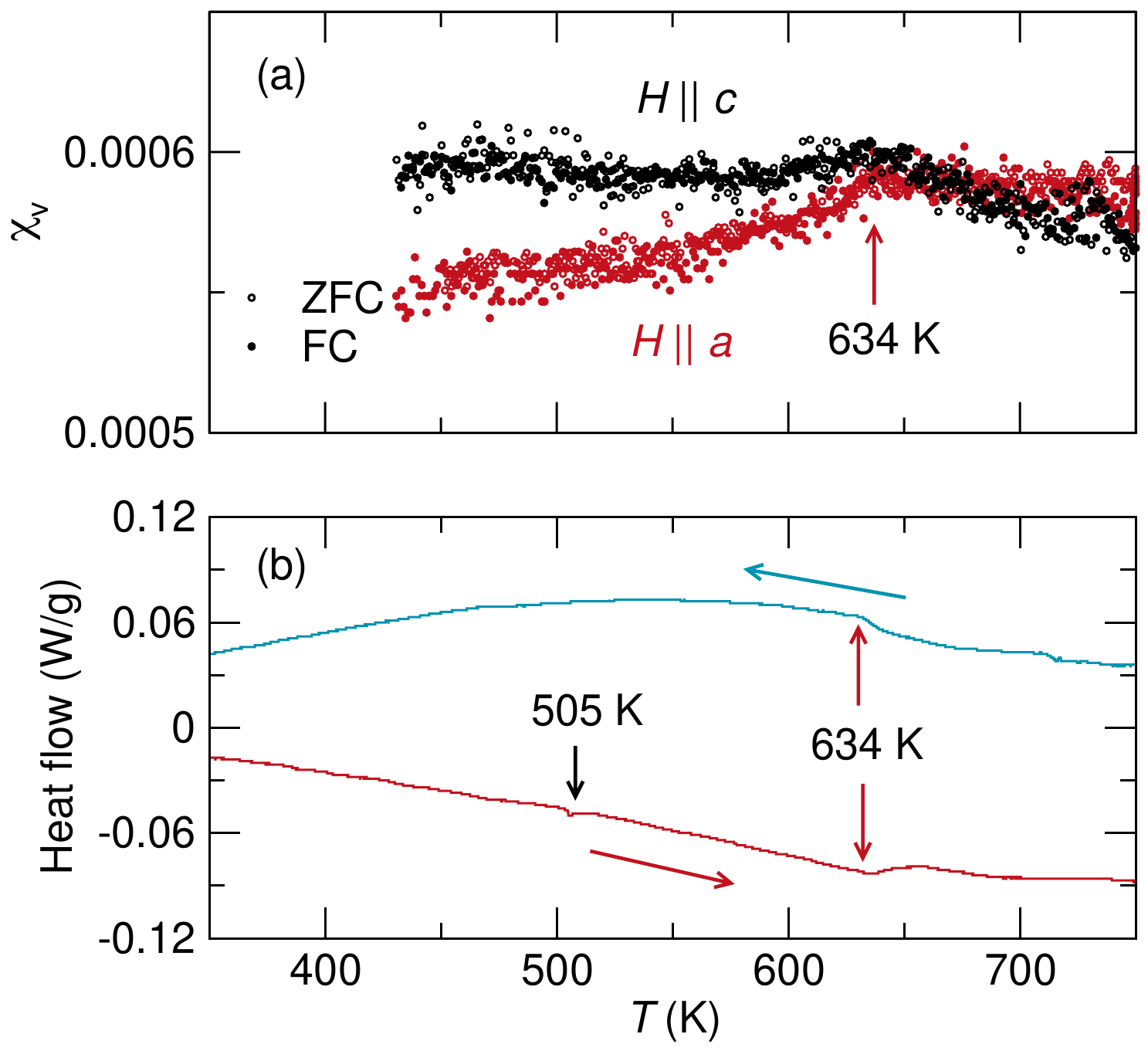} \\
\caption{\label{fig:MvsT}
The temperature dependence of (a) Cr$_2$Al dimensionless magnetic susceptibility $\chi$ under an applied field $H=1~$T, measured to 750~K. Zero-field-cooled (ZFC) and field-cooled (FC) measurements, represented as open and filled circles, overlap in both crystal orientations. In (b), DSC measurements of multiple crystals confirm $T_N$, with a small anomaly at the Sn melting temperature 505 K.
}
\end{figure}

The temperature dependence of the magnetic susceptibility of an aligned single crystal from 430 to 750~K with $H=1$~T is shown in Figure \ref{fig:MvsT}(a). A small hump is observed at $T_N = 634$~K for measurements with $H || a$ and $H || c$. The splitting of the two orientations and the lower susceptibility with field along $a$ indicates that the magnetic ordering is in the $ab$ plane. 
Further confirmation of $T_N$ is seen in calorimetry, shown in Figure \ref{fig:MvsT}(b). The small event at $T=505$~K is the melting of a small amount of residual Sn. \textcolor{black}{ The fraction of residual Sn is estimated to be 2 wt\% based on peak integration of the DSC measurement (see Figure S11).}\cite{supplement,Sn1993}

\begin{figure}
\centering\includegraphics[width=0.9\columnwidth]{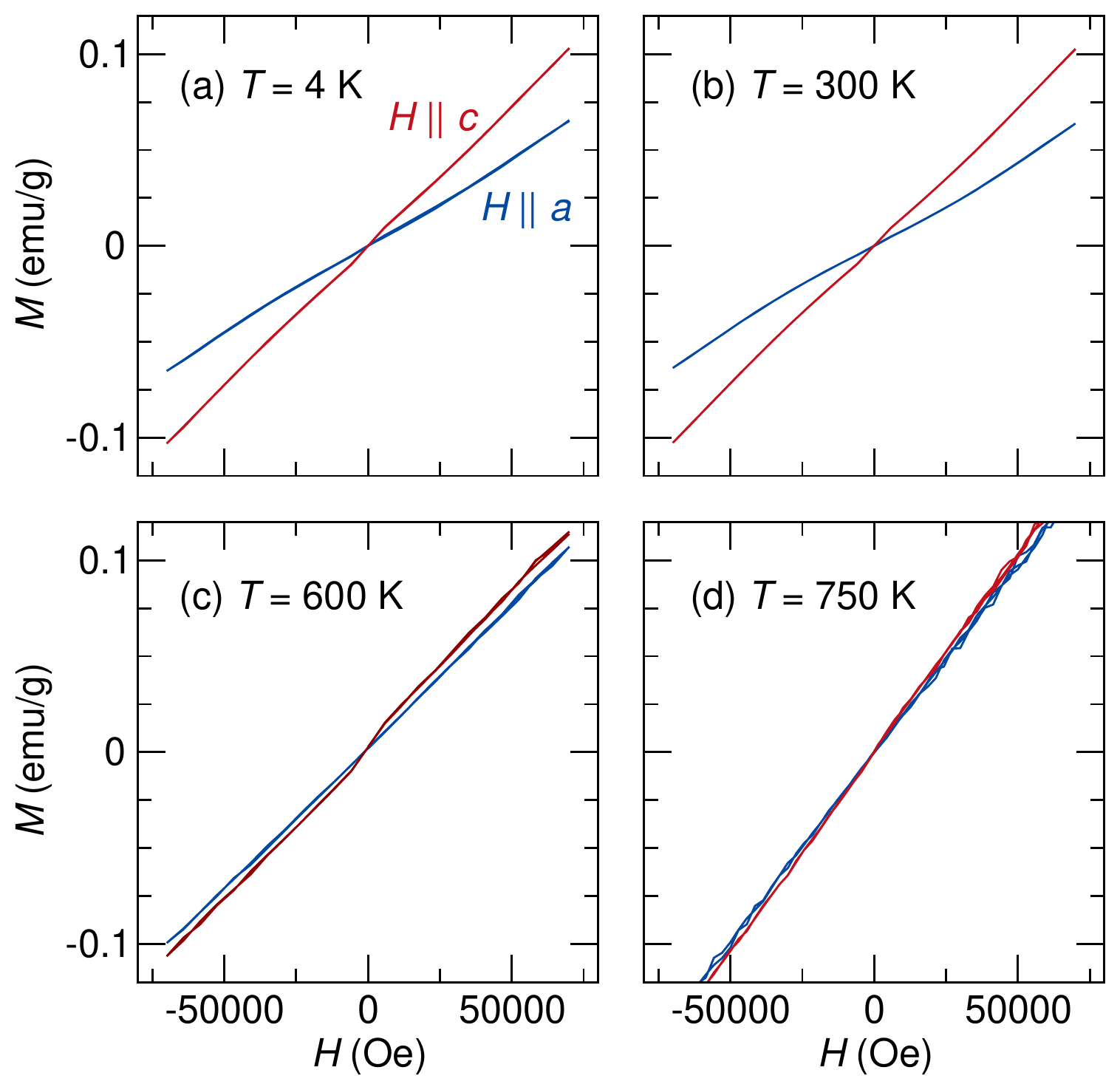} \\
\caption{\label{fig:MvsH}
Field-dependent magnetization of Cr$_2$Al at $T$ = 4, 300, 600, and 750~K (a-d), shown for field along the $a$ (blue) and $c$ axes (red). Anisotropy in the susceptibility vanishes above $T_N \sim 634K$ (a) T = 4 K, (b) T = 300 K, (c) T = 600 K, and (d) T = 750 K, with $H$ along the $a$-axis (blue line) and $H$ along the $c$ axis (red line). 
}
\end{figure}
The magnetization as a function of applied field and orientation is shown in Figure \ref{fig:MvsH}. The anisotropy consistently shows that the susceptibility along $c$ is greater than along $a$, which is to be expected for moments lying in the $ab$ plane. No evidence of a spin flop transition is observed up to $H=7$~T. 
When the temperature of the experiment is raised to 750~K, the anisotropy between $c$ and $a$ directions vanishes, as expected above $T_N$. This is the first measurement of the magnetic anisotropy of a bulk MoSi$_2$-type antiferromagnet in the vicinity of its $T_N$, and understanding the temperature dependence will be a crucial topic of future work.

The phenomenon that the measured moment of Cr$_2$Al is greater with $H || c$ over the entire field and temperature range can be compared to similar measurements on another in-plane degenerate antiferromagnet Fe$_2$As.\cite{Yang}
In Fe$_2$As, the low-field susceptibility mirrors that of Cr$_2$Al. Above approximately 0.7~T, the susceptibility of Fe$_2$As with $H || a$ becomes greater than with $H || c$, indicating rotation of the antiferromagnetic domains into a single-domain state with all moments likely along $b$. This domain rotation is absent in Cr$_2$Al for the fields we are able to achieve here. This lack of domain rotation is most likely due to the small intrinsic susceptibility of Cr$_2$Al versus Fe$_2$As.

The magnetic susceptibility of Cr$_2$Al can be extracted from total energy calculations within DFT for magnetic configurations with tilted magnetic moments. Since an explicit magnetic field is not implemented in our DFT simulations, we use the tilting of magnetic moments to mimic the magnetic structure under an applied external field \cite{Kang:2020}.
Within this approach, we obtain the lowest energy for a given tilting angle. Constraining the tilting angle of magnetic moments changes the total energy changes due to exchange interactions in the tilted state (see Figure S10 \cite{supplement}). We compute the magnetic susceptibility from a quadratic fit to the resulting total-energy curve and the equation 
\begin{equation}
\label{eq:magsus}
\chi_{v}=\frac{\mu_{0}}{2a-\mu_{0}},
\end{equation}
where $\mu_{0}$ is the vacuum permeability and $a$ the quadratic fit coefficient \cite{Kang:2020}. Since DFT implements the calculation at $T=0$~K, $\chi_{a}$ should be zero and $\chi_{c}$ is non-zero for a single crystal with a single domain. Thus, we assume an experiment in which the field is oriented along the $c$ axis and moments are tilted within the $ac$-plane. The calculated magnetic susceptibility $\chi_{c}^{\text{DFT}}$ is 6.88$\times10^{-4}$, which shows good agreement with the scale of measured value of 1.07$\times10^{-4}$ at $T=4$~K, two orders of magnitude smaller than 0.015 for Fe$_2$As.\cite{Yang2020}

\begin{figure}
\centering\includegraphics[width=0.9\columnwidth]{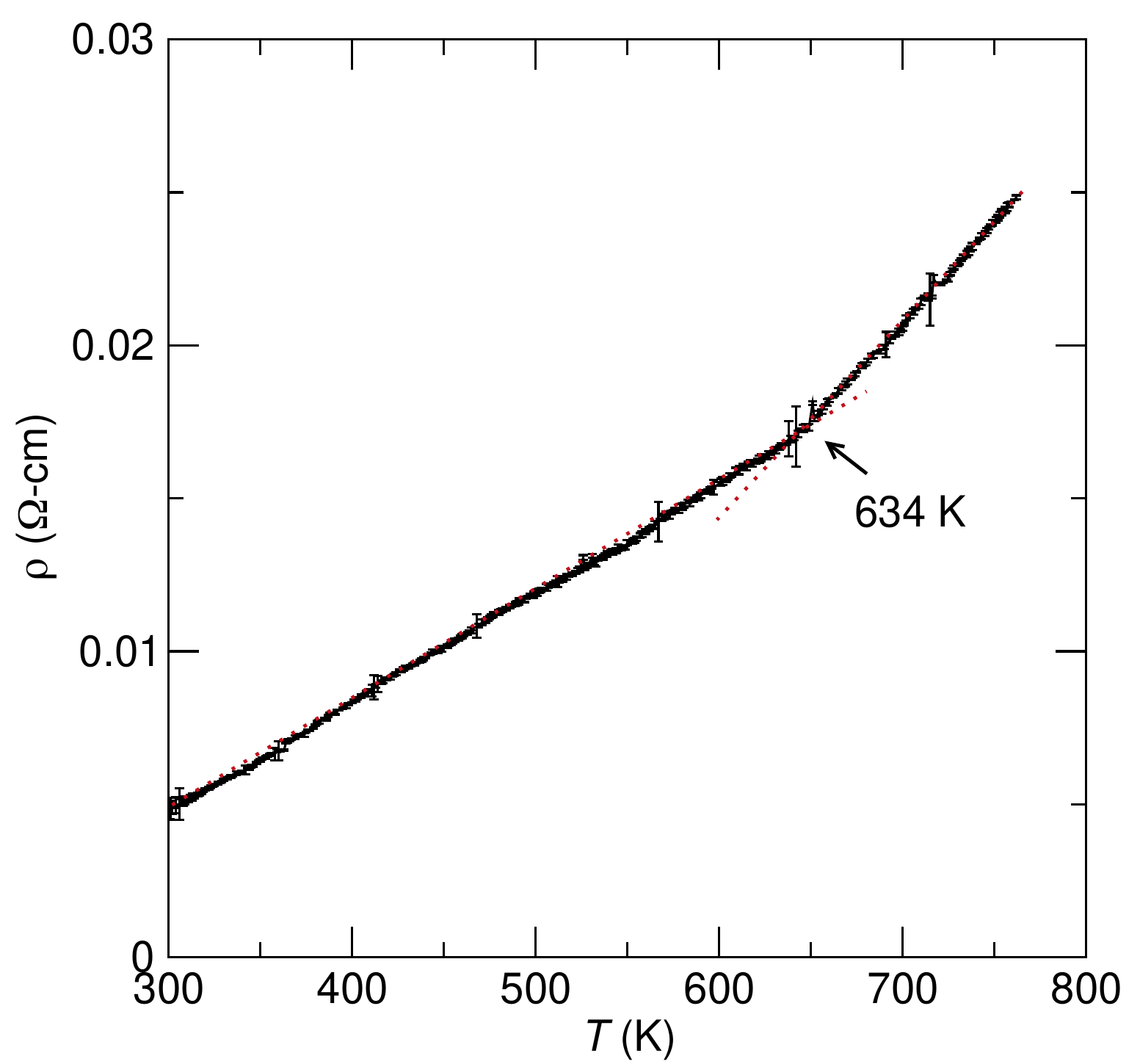} \\
\caption{\label{fig:resistivity}
Two-point resistivity measurement of an unaligned Cr$_2$Al polycrystal shows a clear transition at $T_N$, measured under flowing N$_2$. 
}
\end{figure}

The two-point resistivity of Cr$_2$Al from 298 to 733~K is shown in Figure \ref{fig:resistivity}. Since contact resistance between the Al pads and the Cr$_2$Al sample is present, this data serves as an upper bound for the intrinsic resistivity of Cr$_2$Al.
A clear second-order transition can be observed at 634~K where the slope $d\rho/dT$ increases, indicating an increase in scattering due to the disappearance of AF domains.

\begin{figure}
\centering\includegraphics[width=\columnwidth]{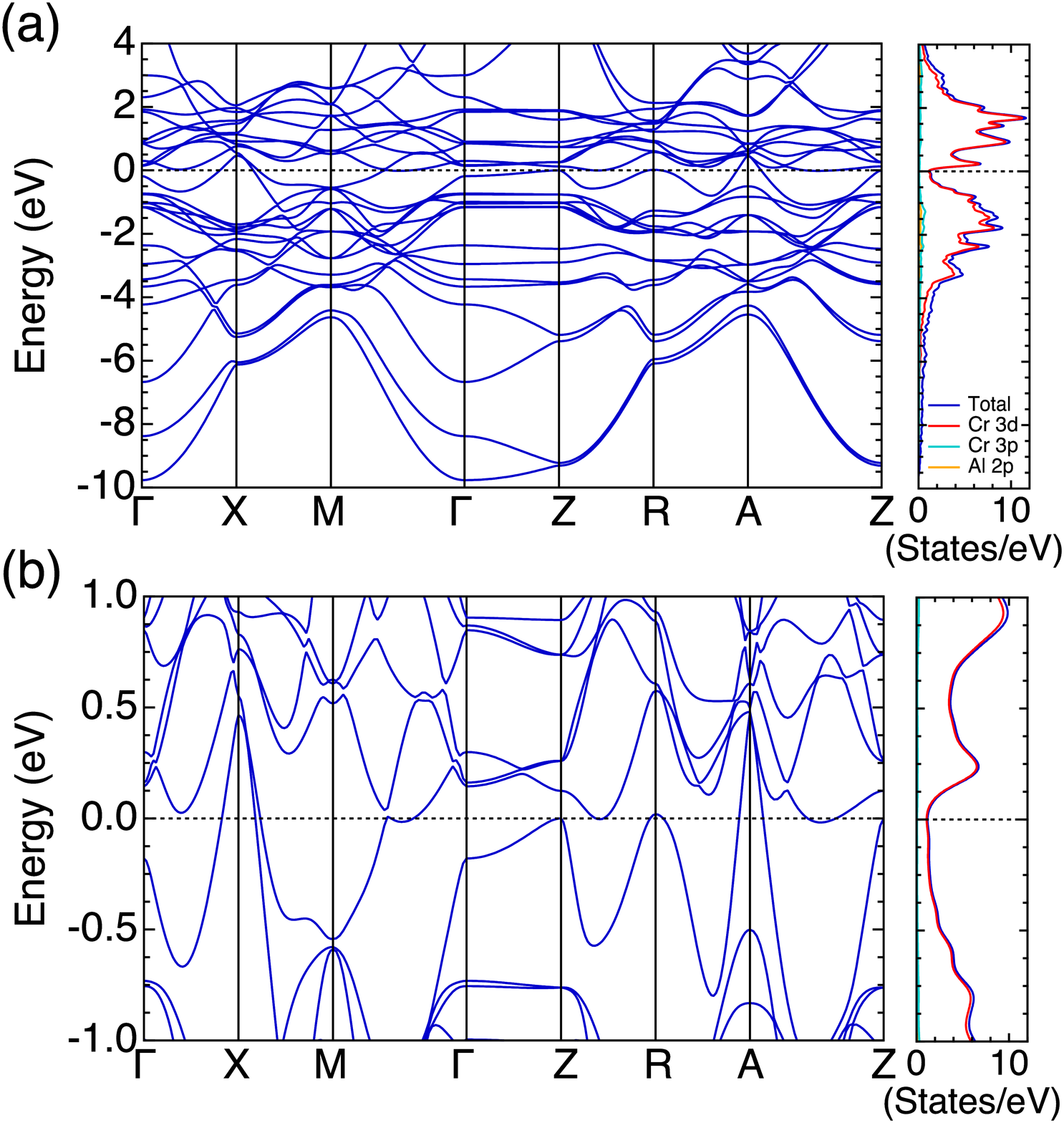} \\
\caption{\label{fig:electronbands}
(a) Electronic band structure and total and projected density of states of Cr$_2$Al and (b) magnified band structure near the Fermi level.
}
\end{figure}

The calculated electronic band structure shown in Figure \ref{fig:electronbands} demonstrates the metallicity of Cr$_2$Al, with a low electronic density of states near the Fermi level, which leads to high resistivity, especially when some antisite defects may be present due to the compositional phase width of Cr$_2$Al.
The number of states at the Fermi level of Cr$_2$Al is 0.0155\,states/(eV$\cdot$\AA),  smaller than that of Fe$_2$As (0.0770\,states/(eV$\cdot$\AA))\cite{Yang2019}. This explains why Cr$_2$Al shows higher resistivity than Fe$_2$As. The projected density of states shows that a wide energy range of states originates from Cr $3d$ orbitals, while $s$ or $p$ orbital contributions from Cr and Al are small.

\begin{figure}
\centering\includegraphics[width=\columnwidth]{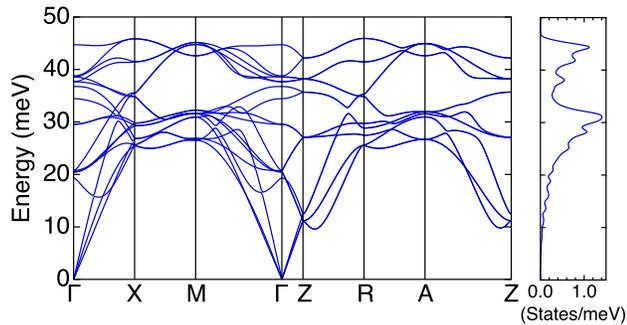} \\
\caption{\label{fig:phononbands}
Phonon bands and density of states of Cr$_2$Al calculated by the finite difference method and DFT.
}
\end{figure}

The structural stability of Cr$_2$Al is confirmed by the phonon dispersion, computed using the finite difference method including the antiferromagnetic configuration and the effect of spin-orbit coupling, shown in Figure \ref{fig:phononbands}. There are no states with imaginary energy/frequency, which confirms the dynamic stability. There are a  total of 18 phonon bands, spanning the energy up to 50\,meV. The phonon density of states shows two peaks around 30 and 45\,meV. 

\begin{figure}
\centering\includegraphics[width=\columnwidth]{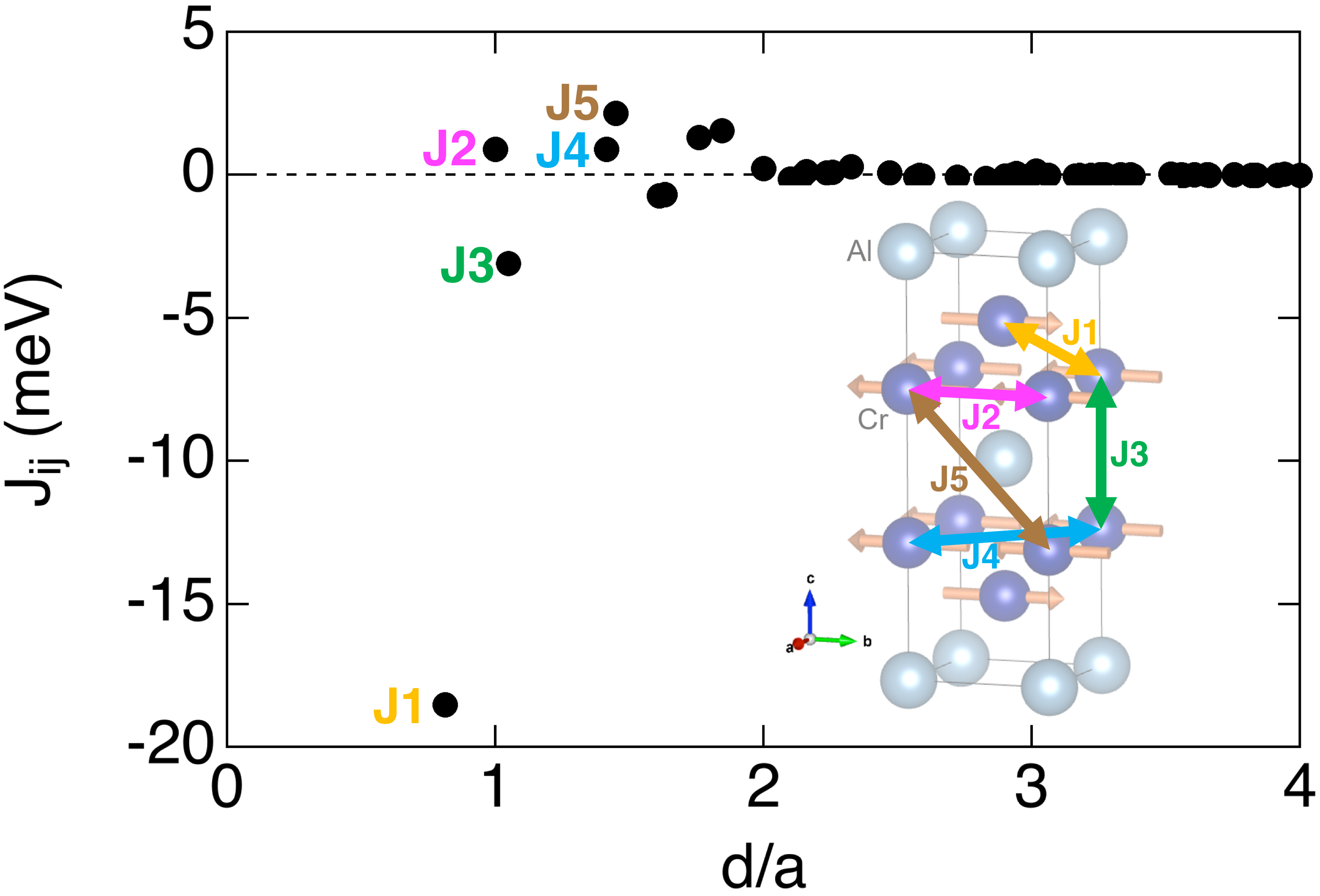} \\
\caption{\label{fig:DFT-exchange}
Exchange coefficients $J_{ij}$ as function of the interaction distance $d$ (in units of the  lattice parameter $a$) between lattice sites $i$ and $j$. First to fifth-nearest neighbor interactions ($J1$, \dots, $J5$) are illustrated in the subset figure with colored arrows.
}
\end{figure}

Exchange coefficients describe the interaction between magnetic moments and can further explain the magnetic structure of Cr$_2$Al. These exchange coefficients can be utilized for atomistic spin dynamics simulations, as has been done for Fe$_2$As.\cite{karigerasi_strongly_2020} The exchange coefficients extracted in this work are shown as a function of relative distance ($d/a$) in Figure\,\ref{fig:DFT-exchange}. Positive and negative exchange parameters indicate ferromagnetic and antiferromagnetic couplings, respectively. {\color{black} The nearest-neighbor exchange interaction ($J1=-18.5$\,meV, shown as a yellow arrow in Figure \ref{fig:DFT-exchange}) is the primary driving force for the AF ordering in Cr$_2$Al. Even though the third-nearest neighbor exchange interaction also shows a negative value ($J3=-3.1$~meV), the corresponding moments show parallel alignment. This is because the $J3$ interaction is overcompensated by $J5=+2.2$~meV, i.e., ferromagnetic coupling, which has a multiplicity four, while $J3$ has only one interaction.
Interactions along $a$ and $b$ directions are all ferromagnetic ($J2=+0.9$~meV and $J4=+0.9$~meV).
Numerical uncertainties of the exchange integrals mainly originate from k-point sampling and we determined these to be lower than 2.7\%, except for J4, where due to the small absolute value the relative error is larger (14.2\%).}

\begin{figure}
\centering\includegraphics[width=\columnwidth]{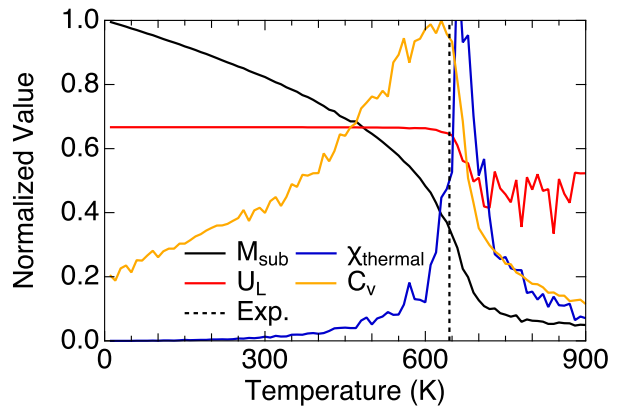} \\
\caption{\label{fig:DFT-neel}
Normalized sublattice magnetization and three thermodynamic observables for Cr$_2$Al, computed from Monte Carlo simulations of the stochastic Landau-Lifshitz-Gilbert equation. The temperature is changed using a step size of 10~K. Heat capacity $C_v$ mirrors the heat flow while heating from DSC in Figure \ref{fig:MvsT}, while the predicted $T_N$ range is 620\,--\,670\,K.
}
\end{figure}

Based on the calculated exchange coefficients, $T_N$ was estimated using three thermodynamic observables that were computed from Monte Carlo (MC) simulations, see details in Ref.\ \onlinecite{Eriksson:2017}.
The observables, computed using the stochastic LLG equation, are shown as a function of $T$ in Figure\,\ref{fig:DFT-neel}. Sub-lattice magnetization ($M_{\text{sub}}$, black solid line) is not sharp enough to determine the critical temperature because of finite size effects, but it does reproduce the 23\% difference between the calculated $T=0$~K $M_{sub}$ and the experimental $M_{sub}$ at 300~K from neutron scattering. Instead, we use the Binder cumulant $U_{\text{L}}$, which indicates $T_N$ as a transition from the normalized value of 0.667 to 0.444. 
The isothermal susceptibility $\chi_{\text{thermal}}$ and the heat capacity $C_{v}$ show a peak at the critical temperature.
This leads to predictions for the transition temperature of 660, 650, and 630~K, respectively, which are all in good agreement with the experimental value of $634 \pm 2$~K. 

\section{Conclusions}

Neutron diffraction and magnetometry are consistent with an in-plane magnetic structure of Cr$_2$Al, identical to that of Mn$_2$Au, but with light elements that minimize spin-orbit coupling. Magnetometry, calorimetry, and transport all confirm  $T_N$ to be $634\pm2$~K, in agreement with previous studies. The ability to access this $T_N$ without decomposing the compound enables a more stringent test for evaluating spintronic phenomena that should vanish in the paramagnetic regime. Aligned measurements provide some early evidence for anisotropy in Cr$_2$Al, which will be the focus of further studies. The ability to engineer the transport and morphology, for example by thin film deposition, should provide a promising platform for spintronic investigation.

\section{Acknowledgments}

This work was undertaken as part of the Illinois Materials Research Science and Engineering Center, supported by the National Science Foundation MRSEC program under NSF Award No. DMR-1720633. The characterization was carried out in part in the Materials Research Laboratory Central Research Facilities, University of Illinois. This work made use of the Illinois Campus Cluster, a computing resource that is operated by the Illinois Campus Cluster Program (ICCP) in conjunction with the National Center for Supercomputing Applications (NCSA) and which is supported by funds from the University of Illinois at Urbana-Champaign. This research is part of the Blue Waters sustained-petascale computing project, which is supported by the National Science Foundation (Awards No. OCI-0725070 and No. ACI-1238993) and the state of Illinois. Blue Waters is a joint effort of the University of Illinois at Urbana-Champaign and its National Center for Supercomputing Applications. This research used resources of the Spallation Neutron Source, a DOE Office of Science User Facility operated by Oak Ridge National Laboratory. The authors thank Jue Liu for additional assistance with the neutron scattering experiment.

\bibliography{Cr2Al_mag}

\end{document}